**Title:** Imaging the ultrafast coherent control of a skyrmion crystal

**Authors:** Phoebe Tengdin[1]†, Benoit Truc[1]†, Alexey Sapozhnik[1]†, Lingyao Kong[2], Nina del Ser[3], Simone Gargiulo[1], Ivan Madan[1], Thomas Schönenberger[4], Priya R. Baral[5], Ping Che[6], Arnaud Magrez[5], Dirk Grundler[6], Henrik M. Rønnow[4], Thomas Lagrange[1], Jiadong Zang[3,7] Achim Rosch[3], Fabrizio Carbone*[1]

*fabrizio.carbone@epfl.ch
†these authors contributed equally

**Affiliations** :
[1]Institute of Physics, LUMES, École Polytechnique Fédérale de Lausanne (EPFL), Lausanne, Switzerland
[2]School of Physics and Optoelectronics Engineering Science, Anhui University, Hefei 230601, China
[3]Institute for Theoretical Physics, University of Cologne, Köln, Germany
[4]Institute of Physics, LQM, École Polytechnique Fédérale de Lausanne (EPFL), Lausanne, Switzerland
[5]Institute of Physics, Crystal Growth Facility, Ecole Polytechnique Fédérale de Lausanne (EPFL), Lausanne, Switzerland
[6]Institute of Materials (IMX), Laboratory of Nanoscale Magnetic Materials and Magnonics, Ecole Polytechnique Fédérale de Lausanne (EPFL), Lausanne, Switzerland
[7]Department of Physics and Astronomy, University of New Hampshire, Durham NH, USA

**Abstract:** Exotic magnetic textures emerging from the subtle interplay between thermodynamic and topological fluctuation have attracted intense interest due to their potential applications in spintronic devices. Recent advances in electron microscopy have enabled the imaging of random photo-generated individual skyrmions. However, their deterministic and dynamical manipulation is hampered by the chaotic nature of such fluctuations and the intrinsically irreversible switching between different minima in the magnetic energy landscape. Here, we demonstrate a method to coherently control the rotation of a skyrmion crystal by discrete amounts at speeds which are much faster than previously observed. By employing circularly polarized femtosecond laser pulses with an energy below the bandgap of the Mott insulator $Cu_2OSeO_3$, we excite a collective magnon mode via the inverse Faraday effect. This triggers coherent magnetic oscillations that directly control the rotation of a skyrmion crystal imaged by cryo-Lorentz Transmission Electron Microscopy. The manipulation of topological order via ultrafast laser pulses shown here can be used to engineer fast spin-based logical devices.

## I. INTRODUCTION

When an electron traverses a skyrmion's magnetic structure, the topological ordering causes the electron's spin to pick up a Berry phase. This causes a Lorentz force on the electron as well as a net force on the skyrmion oriented perpendicular to the flow of electric current, known as the Skyrmion Hall effect [1], [2]. The effect provides a greatly enhanced coupling of electric current to the magnetic texture, much more efficient than for current driven manipulation of domain walls [3], [4]. In analogy to the case of electric current, skyrmions present in an insulating host material will be subject to similar forces when exposed to a pure spin current [5]. However, this process can proceed without the Ohmic losses that exist when using electrical current. Additionally, excitation of spins can be achieved in an ultrafast and contact free manner using ultrafast lasers on femtosecond timescales [6], [7].

The emergence of $Cu_2OSeO_3$ as a skyrmion hosting Mott insulating material with multiferroic properties and bulk Dzyaloshinskii-Moriya Interaction (DMI) opens the possibility to study and manipulate topological order and skyrmion dynamics purely under the influence of magnetic excitations or electric fields[8]. Additionally, spin currents and collective oscillations in $Cu_2OSeO_3$ have been shown to have an exceptionally low damping and correspondingly long mean free path, making them effective candidates for manipulating spin order [9], [10]. Recent works have demonstrated the ability to rotate the skyrmion crystal in $Cu_2OSeO_3$ via thermally generated spin currents [11], [12], electric fields [13], [14], and a magnetic field gradient in doped crystals [15]. In [11], [12], the spin currents were induced via the Spin Seebeck effect with a strong local heat gradient generated from a high power electron beam. For all previous experiments of this kind, the rotation proceeded on the timescale of hundreds of milliseconds to seconds. To increase the speed of these processes, faster excitation mechanisms are required.

Recent experiments have shown that circularly polarized femtosecond pulses of light can induce an effective magnetic field of up to 0.6 Tesla in a material for timescales as short as 50-100 femtoseconds and drive switching of the magnetic order via the inverse Faraday effect [16]–[18]. Femtosecond light pulses can also generate spin excitations that have pulse widths in the fs timescale and can travel up to ballistic speeds [19]–[23]. However, the microscopic details of spin excitation on ultrafast timescales are not fully understood. Experiments investigating these excitations have been ultimately constrained by a limited ability to directly image spins on the relevant length (nanometer) and time (femtosecond) scales and have furthermore been limited by the inherently irreversible nature of many of the magnetic phenomena studied. The development of Lorentz force microscopy for imaging magnetic textures such as skyrmions coupled with ultrafast excitation of these structures constitute promising tools for enhancing the understanding of these spin excitations and their propagation.

In this work, we take advantage of the strong coupling between topological ordering and collective spin oscillations in $Cu_2OSeO_3$ to drive skyrmion rotational motion with single femtosecond pulses of circularly polarized near-infrared light. After each individual laser pulse, we image the skyrmion crystal in real space via *in-situ* cryo-Lorentz force transmission electron microscopy (L-TEM). We show that with each laser pulse we rotate the skyrmion crystal by a controlled and irreversible amount. The magnitude of the rotation depends sensitively upon the polarization and fluence of the pulse. With time-varied double pulse measurements performed at the fluence threshold of the observed rotation, we show that this rotation process is driven by a collective magnon excitation that has a characteristic excitation period of ~175 picoseconds. Furthermore, the rotation can be switched on and off in a coherent manner by changing the delay time between successive driving pulses with the appropriate polarization. Our conceptually new experimental protocol provides nanoscale images of irreversible modification of the skyrmion crystal orientation that correspond to picosecond dynamics in the material. Through real space analysis of our images, we generate detailed mappings of the rotations present over macroscopic distances (10s of microns) with a precision that is only limited by the natural length scale of the skyrmions themselves (~60nm). Additionally, the energy of the light used for excitation (1 eV) is far beneath the bandgap of the skyrmion host material, and thus control can be achieved with remarkably low values of absorbed fluence, potentially enabling future ultrafast and highly efficient devices.

## II. EXPERIMENTAL RESULTS

Figure 1(a) and (b) show real space cryo-LTEM under-focused images of the skyrmion crystal in $Cu_2OSeO_3$ discussed in this work. Figure 1(a) shows a metastable skyrmion crystal that forms when we cool the thin lamella from above the Curie temperature (~60K) to 5K under an applied magnetic field of 34mT. Next, we irradiate the sample with individual femtosecond laser pulses, and observe that the skyrmion crystal rotates. Figure 1(c) illustrates our procedure for tracking the rotation of the skyrmion crystal in the real space TEM images. After each successive laser pulse, we take an image and then take the Fourier transform (FT) of the real-space image (or a subsection of an image) and calculate the angle of the FT in a polar coordinate system. We repeat this process, allowing us to map the change in the angle of the skyrmion crystal following a train of pulses of near-infrared radiation. Further details about the pulse train and imaging settings are given in the Methods section. In Fig. 1(d), we extract and plot the angle of a single peak in the FT of the skyrmion crystal after illuminating the sample with femtosecond pulses of light. For the circularly polarized light (both $\sigma^+$ and $\sigma^-$), each pulse rotates the skyrmion crystal by a discrete amount, with the direction of rotation being the same for both handedness' of polarization, while the linearly polarized light does not rotate the skyrmion crystal. This difference between the rotation of linear and circular polarizations implies that the circularly polarized pulses can drive excitation of magnons (the quanta of spin current) on ultrafast timescales. The mechanism for this process is known [24], and is discussed later in the text.

In Fig. 2(a) we plot the fluence dependence of the rotation process. We observe that the magnitude of rotation depends sensitively on the amount of laser fluence used in the experiment. In Fig. 2(b) we show that the threshold fluence needed to rotate the skyrmion crystal with circularly polarized light is >1.6 mJ/cm$^2$. Above this threshold the rotation amount proceeds in a roughly linear fashion until 8 mJ/cm$^2$. We observed that pulses with energies above this value can melt and reform the skyrmion crystal, thus they cannot be used to rotate the crystal in a controlled way.

Following this fluence dependence, we study the threshold between rotation and nonrotation using a time resolved technique. We split the photoexcitation pulse into two parts with equal value and we measure the rotation of the skyrmion crystal as a function of the time delay between the two pulses, each with a fluence of 1.1 mJ/cm$^2$, corresponding to half of the required fluence for the skyrmion crystal rotation. When the pulses are combined into one pulse at time zero, the excitation is above the threshold where rotation occurs (see Figure 2a). In Fig. 2(c), we plot the difference in the Fourier transform of two images: one image taken before and one taken after two pulses were sent at the intervals indicated (0-300 ps). We observe that rotation occurs only when the pulses are sent at certain intervals.

This result is further investigated in Fig. 3. We plot the detailed time dependence of the skyrmion crystal rotation as function of pulse separation in Fig. 3(a). The blue data points were taken for a sequence of two right-handed circular ($\sigma^+$) polarized pulses. The time dependence of the rotation phenomena shows an oscillation with a period of ~175 picoseconds and a damping that takes place over the course of a nanosecond. This response can only be attributed to the launching of a coherent collective magnetic oscillation in the skyrmion crystal that drives the rotation process. For the red data points, we send one pulse with $\sigma^+$ polarization and a second pulse with left-handed circular polarization ($\sigma^-$) and slightly more than half the fluence value of the first pulse (0.6 mJ/cm$^2$). For a sequence of $\sigma^+ + \sigma^-$ pulses, we observe a coherent drive that is out of phase with the $\sigma^+ + \sigma^+$ sequence by 180 degrees (~87 ps). This is due to the polarization of the second pulse, which can excite a magnetic field with opposite direction to the first one via the inverse Faraday effect. The magnitude of the rotation also has a weaker amplitude due to the weaker amplitude of the second pulse, showing that the process roughly scales linearly as predicted in Fig. 2(a-b).

Next, we present a theoretical model that can help us to understand the origin of the collective excitation observed. We compute and show theoretically in the supplementary information that the combination of Gilbert damping and breathing-mode oscillations naturally leads to rotational torques, $T_{\alpha,\text{pump}}^R = -\alpha m_0 \int d^3\vec{r} \, \frac{d\hat{n}}{d\theta} \partial_t \hat{n} \propto (\delta M)^2$ , where $m_0$ is the spin density, $\alpha$ the Gilbert damping, $\hat{n}$ the direction of the magnetization and $\frac{d\hat{n}}{d\theta}$ the change of the magnetization as function of the rotation angle $\theta$. For a clean system we compute the rotation angle $\Delta\theta_{\alpha,\text{pump}}$ after a field pulse of amplitude $\delta B$ and duration $\tau$ and obtain (see supplementary material)

$$\Delta\theta_{\alpha,\text{pump}} \approx \gamma_{\alpha,\text{pump}} \frac{1}{\alpha\, N_S}\left(\frac{\delta B}{B_0}\right)^2 \left(\frac{\tau}{T_0}\right)^2. \qquad (1)$$

where $T_0$ is period of the breathing-mode, $B_0$ the static external magnetic field, $N_S$ the number of skyrmions involved in the rotation and $\gamma_{\alpha,\text{pump}} \approx 9.7°$ is a prefactor for a single pulse which we determined using numerical simulations, see SI. For a sequence of two pulses with relative delay time $\Delta\tau$ we show in Fig. 3(b) how the rotation angle changes as function of $\Delta\tau$. This qualitatively reproduces the experimental result of Fig. 3(a). The remarkable match in the timescales of the experimental and theoretical data confirms that rotational torques induced by the breathing mode can explain our experiment.

Further analysis of our system shows that all rotations are well completed within 5 ns. If we assume a rotation timescale of 5-50 ns, we estimate an effective rotation rate of $2\text{x}10^8$ – $2\text{x}10^7$ deg/sec, see SI. This rotation rate is more than six orders of magnitude faster than previously reported [11]. If we consider the case of coherent control, when the oscillations are stopped after a half period of 87 ps by a pulse of the appropriate amplitude, the effective rotation rate could even be increased to $2\text{x}10^{10}$.

In Fig. 4 we use our real space image to map the rotation of the skyrmion crystal across macroscopic distances while again illuminating the sample with a train of femtosecond near infrared (1 eV) laser pulses. We divide the real space image of Fig. 4(a) into 25 individual boxes, for each of which we take the FT of the (sub)image and then plot the region around a single point in the FT. In Fig. 4(b-d) we show cutouts of the rotation that takes place in different regions of the sample. See SI for additional plots of the rotation present in each subregion of a 5x5 grid of the sample. We found that by analyzing subregions of the samples, we could better isolate the different regions of the skyrmion rotation, resulting in a higher quality signal. For the data presented in Fig. 4, the grid size that led to the best isolation of rotation regions corresponded to an analysis subregion of 460 nm. Since the skyrmion crystal period in this material is approximately 60 nm, this corresponds to a cluster of 7-8 skyrmions across, or approximately 50 skyrmions. We observe rotations as low as 0.05 degrees per pulse in the Fourier transform of our images, here taken after a sequence of 120 pulses, this corresponds to movements of the skyrmions in real space of about 0.4 nanometers (per pulse).

## III. DISCUSSION

Several mechanisms have previously been identified as potential origins for the rotation of skyrmion crystals. Roughly, they can be grouped into two different classes. The first one is based on a manipulation of anisotropy terms, e.g., by electric fields [13], [14], which leads to a rotation by a finite angle. In contrast, a continuous rotation can be induced using the Magnus force imprinted onto the skyrmions by electrical, spin- or heat currents. In an inhomogeneous system these forces lead to a rotational torque. The inhomogeneity can arise from, e.g., a temperature gradient [4] or simply radial heat currents when the center of the skyrmion crystal is heated[11], [12]. While the latter effect may be of relevance in our experiment at high fluence density, the polarization- and time-dependence of our two-pulse low fluence experiments in Fig. 3(a) allow us to identify uniquely a new mechanism for the skyrmion rotation.

As shown in Fig. 3(a), the first pulse induces a collective oscillation of the skyrmion crystal with a period of T=175 ps corresponding to a frequency of 5.7 GHz. The origin of these oscillations is already known[24] [16]: via the inverse Faraday effect, the polarized laser light induces an effective magnetic-field pulse which triggers the breathing mode of the skyrmion crystal, i.e., a coherent oscillation of the size of each skyrmion. The magnetic-field pulse has an amplitude of ±14 mT for left- or right-polarized light with a duration of $\tau = 50$ fs, [16], [25], see SI for details. The coherent oscillations are enhanced if the second pulse is either in-phase with the same polarization as the first pulse or out-of-phase with the opposite polarization, as seen for the red data points. Remarkably, the skyrmion rotation starts whenever the amplitude of the breathing mode oscillation is sufficiently large. This is direct experimental evidence that collective oscillations induced by the inverse Faraday effect lead to rotations of the skyrmion crystal.

Previous numerical and analytical studies[26], [27] have shown that collective magnetic oscillations can induce a translational motion of magnetic textures proportional to $\delta M^2$, where $\delta M$ is the amplitude of the oscillatory mode. In the current setting such a translational motion is prohibited by symmetry. Quantitatively, however, Eq. (1) predicts – for a clean system and using the ultrashort pulse duration of our experiment – rotation angles about 6 orders of magnitude smaller than observed experimentally, see SI. This shows that the Gilbert damping is probably *not* the primary source of the rotational torques. Instead, the rotational torques are most likely induced because the breathing mode triggers a ratchet-like motion in the disorder potential of our sample. Disorder is furthermore responsible for the fact that the rotation angle is not proportional to $\delta B^2$, instead following the typical threshold behaviour, see Fig. 2(b), expected for a disorder-pinned system. As the system is in a regime where pinning effects are much more important than effects from the very weak Gilbert damping, it is not surprising that disorder leads to substantially larger rotational torques than predicted for a clean system. This is also consistent with the numerical observation in [28] that boundaries can strongly enhance the ratchet-like motion of skyrmions in oscillating magnetic fields.

In conclusion, our experiment shows that single ultrafast laser pulses can trigger remarkably large rotations of skyrmions using rotational torques induced by collective spin oscillations and a ratchet-like motion in a disordered system. We show that fluence, polarization, and timing can directly control the rotation, while other parameters such as the beam shape have not yet been explored. To modify the spin currents and operate spin-based devices, we could imagine using spatially varied laser beam profiles such as Laguerre-Gaussian beams to generate tailored device frameworks as needed for logical operations. Tightly focusing these beams may also offer the possibility to generate individual skyrmions, as shown in ref. [29], while the orbital angular momentum in the beams may lead to even more efficient rotations [30], [31]. Thus, our work offers the possibility to design new modifiable spintronic devices with logical bit sizes limited only by the spatial pattern of the light used for excitation, and with temporal command sequences that can be modified on picosecond timescales. This demonstration of picosecond control over nanometer scaled topological magnetic objects will lead to an array of new device physics and allow scientists to build new functionalities for skyrmions.

## ACKNOWLEDGEMENTS

We acknowledge useful discussions with Ido Kaminer. We acknowledge support from the ERC consolidator grant ISCQuM and SNSF via sinergia nanoskyrmionics grant 171003, the Humboldt Foundation, the DFG via SPP 2137 (project number 403505545) and CRC 1238 (project number 277146847, subproject C04), U.S. Department of Energy, Office of Basic Energy Sciences under grant No. DE-SC0020221, the National Natural Science Foundation of China under grant No.11974021, and the SMART-electron project that has received funding from the European Union's Horizon 2020 Research and Innovation Program under Grant Agreement 964591. S.G. acknowledges support from Google Inc.

## APPENDIX A: METHODS AND MATERIALS

### 1. Methods: Details of the experimental setup

The experiments were carried out in a modified JEOL JEM2100 TEM [32]. In this instrument, *in-situ* cryo-LTEM can be performed in the Fresnel configuration [33] at camera-rate temporal resolution (ms) using a continuous wave electron beam generated thermionically, upon in situ pulsed optical excitation of the specimen with tunable fs source. The camera used for the detection of the electrons was a Gatan® K2 direct detection camera. The sample was cooled to 5K using a helium-cooled sample holder from Gatan.

A Ti:Sapphire regenerative amplifier was used to generate 35-fs pulses of light with a center wavelength at 800 nm and a 34-nm (FWHM) bandwidth. The pulse energy directly from the amplifier was 1.5 mJ per pulse at a 4-kHz repetition rate, and ~55% of this light (0.81 mJ) was used to convert to near-IR wavelength via an optical parametric amplifier (OPA). After conversion to 1200nm/1eV, the pulses have the duration of 50 fs, and we used a series of optical choppers to lower the repetition rate to 10 Hz. In this way, we were able to use a mechanical shutter to send individual pulses as desired, or to send a train of pulses that had a repetition rate lower than the exposure time of our camera. For the pulse train measurements, the pulses had a repetition rate of either 10 Hz or 10 Hz with every 3rd pulse missing (to check for stability). The camera rate was 20 Hz. For the time resolved measurements, pairs of pulses were either sent individually or at 4 Hz repetition rate, with total rotation recorded after 120 pulses and the rotation per pulse calculated by dividing the observed rotation angle by 120. The magnetic field in the microscope was applied normal to the sample surface along the [111] direction.

## 2. Materials: Sample preparation

A high-quality single crystal of $Cu_2OseO_3$ was grown by the chemical vapor transport method. 25g of a stoichiometric mixture of CuO and $SeO_2$ are sealed in a 36mm diameter quartz ampule together with 100mbar of HCl used as transport agent. The ampule is placed in a horizontal two-zones furnace. During the growth, source and sink temperatures are set at 635°C and 545°C respectively. The single crystal was aligned and cut into a cube so that the three main directions correspond to [$1\bar{1}0$], [111] and [$\bar{1}\bar{1}2$], respectively. Then, choosing [111] as the main surface, the cube was cut into slices of ≈ 0.5 mm thickness. The sample was thinned to about 110 nm by Focused Ion Beam (FIB) technique.

## FIGURES

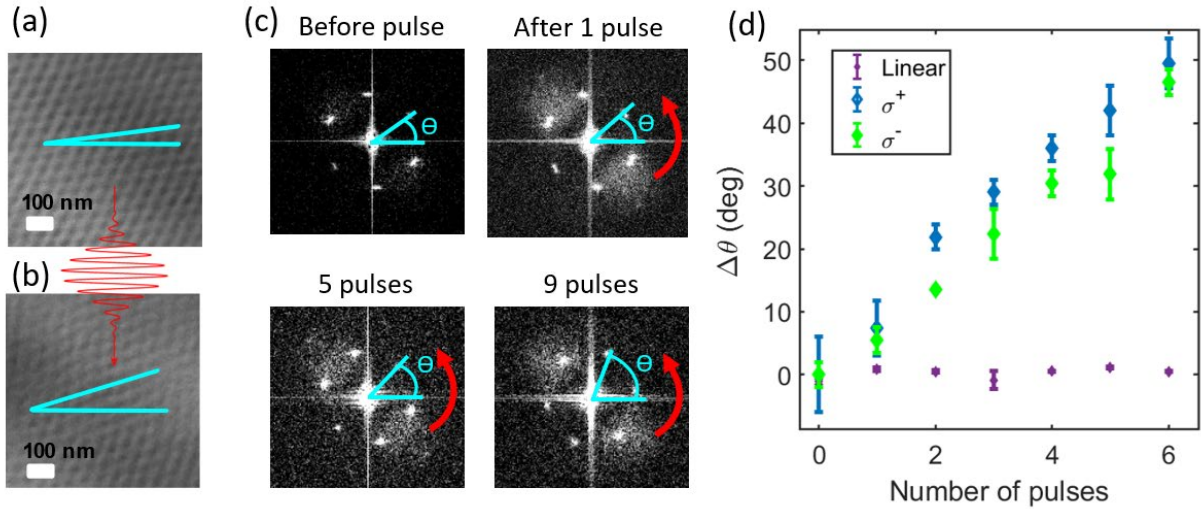

FIG. 1. Illustration and schematic of laser-driven skyrmion crystal rotation process. (a) Real space images of skyrmion crystal in $Cu_2OseO_3$ taken before excitation with the laser pulse and

(b) after 6 successive near-infrared laser pulses have each rotated the skyrmion crystal by a discrete amount. Note that the angle (depicted in blue) to the horizontal has changed. (c) Fourier transforms of L-TEM images following successive pulses of near-infrared femtosecond laser excitation. The angle of the hexagonal ordering of the skyrmion crystal changes as a function of the number of pulses applied to the sample. (d) Tracking the position of a single peak in the FT of an image while pumping the sample with individual femtosecond laser pulses. Note that $\sigma^+$ and $\sigma^-$ polarization both rotate the skyrmion crystal in the same direction, while linear polarization does not rotate the skyrmion crystal. The error bars are 95% confidence intervals calculated from the data and multiplied by an uncertainty factor determined from the noise level in the data.

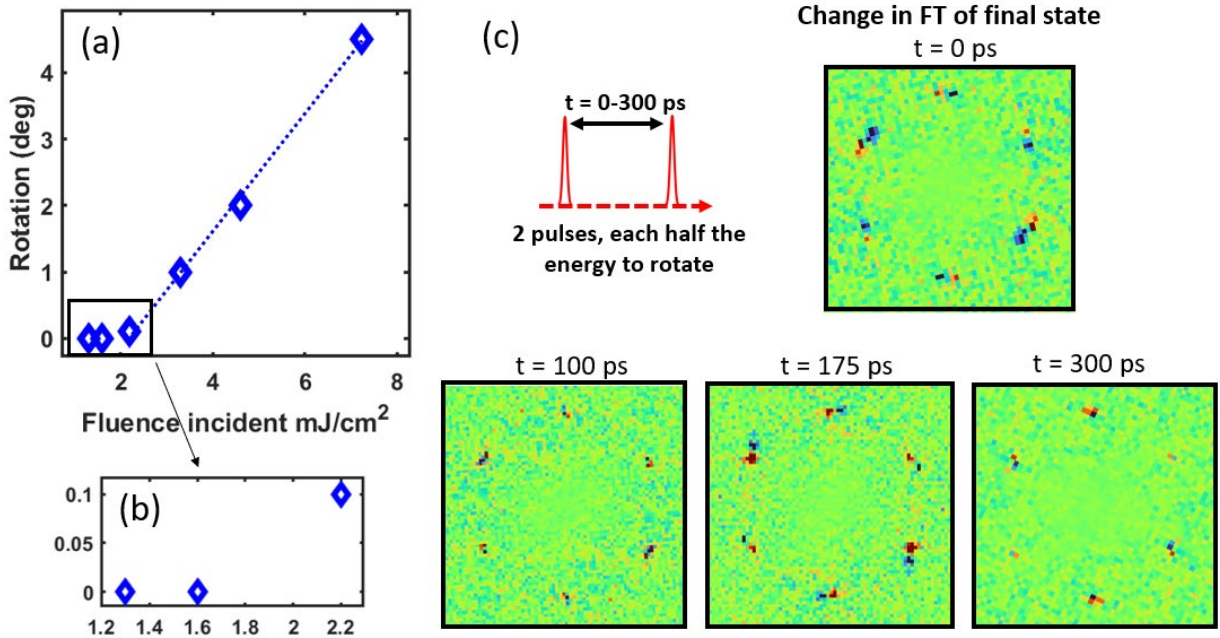

FIG. 2. Fluence and time dependence of skyrmion crystal rotation. (a) Fluence dependence of skyrmion crystal rotation. The dashed line is a linear fit to the data points with fluences from 1.6 mJ/cm$^2$ to 8 mJ/cm$^2$. (b) Threshold value of fluence needed to drive rotation in the skyrmion crystal. This value of fluence (2.2 mJ/cm$^2$) was split into two pulses and used to perform the double pulse time resolved experiments described in (d). Two laser pulses, each with a fluence of 1.1 mJ/cm$^2$ are separated by a controlled delay. After the pulses have excited the sample, a Lorentz image of the magnetization is recorded. The change in the Fourier transform of the images is shown for various time delays between pulses, illustrating that this time delay between the two pulses directly influences the observed rotation. For clarity, the observed changes are shown after the rotation has been driven by 120 pairs of pulses.

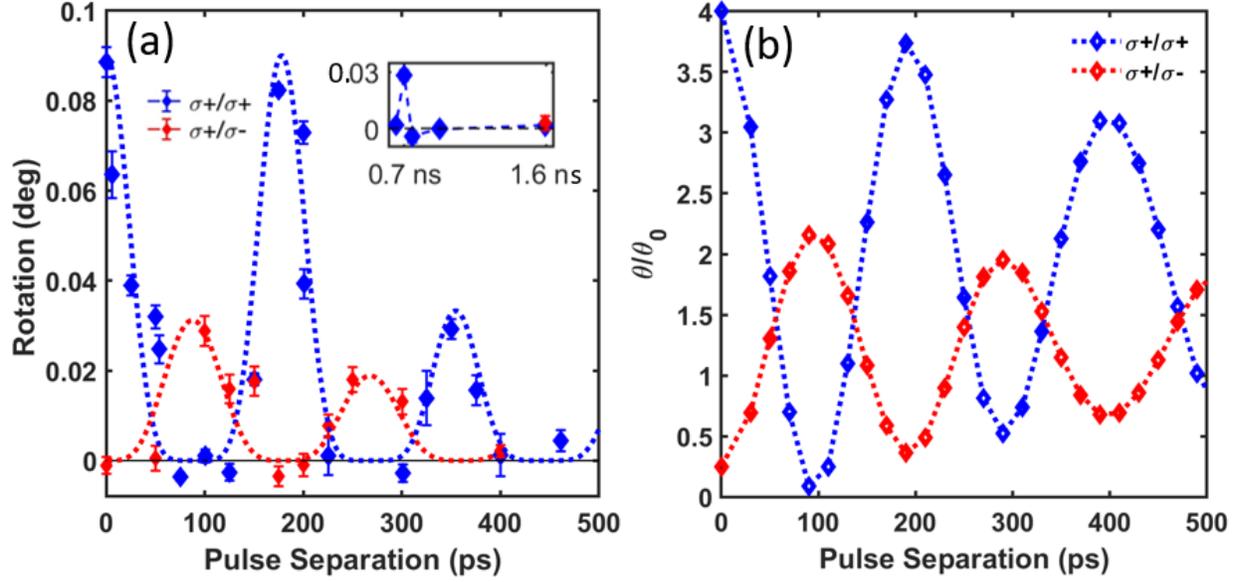

FIG 3. Detailed time dependence of skyrmion crystal rotation. (a) Double pulse timed experiments showing the rotation of the skyrmion crystal observed as function of delay between the two pump pulses. The pulses either have both $\sigma^+$ polarization (blue points) or a sequence of $\sigma^+/\sigma^-$ polarization (red points). For the blue points we used pulses of equal amplitude and observe a coherent oscillation in the amplitude of rotation with a period of 175 ps that damps out progressively over a nanosecond. For the case of $\sigma^+/\sigma^-$ polarization, the second pulse of $\sigma^-$ polarization has half the amplitude of the first one (with $\sigma^+$ polarization). Here we observe rotation with approximately half the amplitude out of phase with the previous oscillation. The blue/red dashed lines are a guide for the eye. Error bars (95% confidence intervals) are within the markers used. (b) Theoretical prediction of the rotation angle $\theta/\theta_0$ for such pulse sequences in a clean system where $\theta_0$ is the rotation angle for a single pulse. The theory is based on the calculation of rotational torques arising from breathing-mode oscillations, see SI.

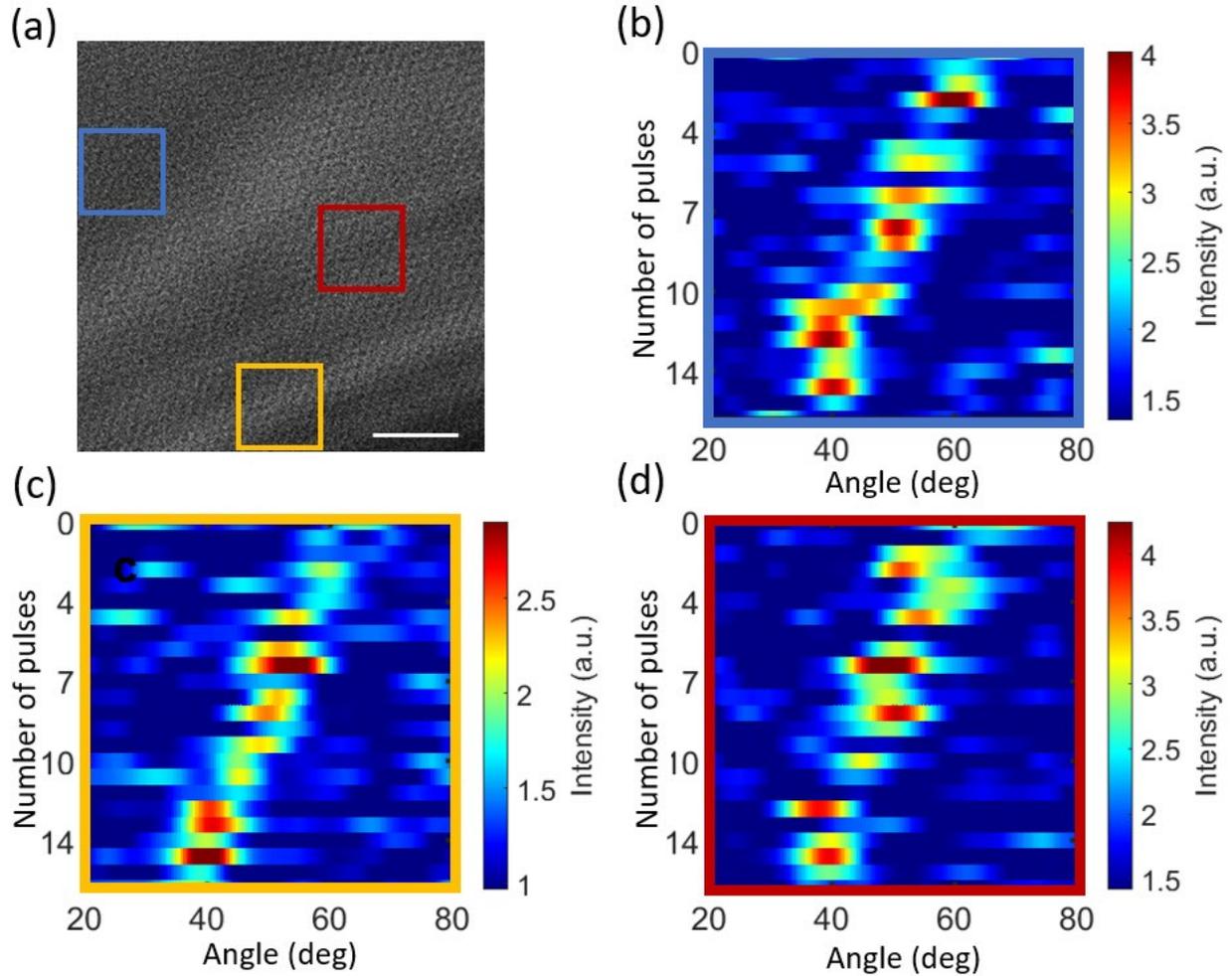

FIG 4. Skyrmion crystal rotation in a real space image from the TEM. (a) Real space L-TEM image of the magnetic structure in $Cu_2OseO_3$. Scale bar (bottom right) is 500 nm. Response of the skyrmion crystal after excitation by a femtosecond laser pulse train is shown for specific regions of the film in (b)-(d). We take the FT of each subsection of the image and plot the angle of a single peak in the FT as a function of the number of pulses applied to the sample. The intensity corresponds to the intensity of the peak in the Fourier transform within a region of angles. See SI for rotational maps of the entire film.

**Author contributions:**

Conceptualization: PT,BT,AS,FC

Data analysis: BT

Experimental methodology: PT,BT,AS,IM,TL

Experimental Investigation: PT,BT,AS

Visualization: PT,BT,AS,TL

Sample preparation: TS,PB,PC,AM

Theory and Simulations: LK,NdS,SG,JZ,AR

Supervision: AM,DG,JZ,AR,TL,FC

Writing-original draft: PT

Writing- review and editing: PT,BT,AS,SG,PC,AM,DG,HK,TL,JZ,AR,FC

**Competing interests:** The authors declare no competing interests.

**Data and materials availability:** All data, code, and materials used in the analyses is available to readers on request.

**Supplementary Media**

1.  SchematicAnimation_2pulse_addition.mov
    Description: Schematic movie of skyrmion breathing mode and rotation dynamics following two pulses with a temporal separation of 175 ps.

2.  SchematicAnimation _2pulse_subtraction.mov
    Description: Schematic movie of skyrmion breathing mode rotation following two pulses with a temporal separation of 88 ps.

3.  SkyrmionModeOf4.8GHzAnd5GHz.mov
    Description: Movie showing the skyrmion breathing modes excited by a magnetic field pulse out of the plane in a film of finite thickness.

**Supplementary Text:**

**Contents:**

## 1. Theory of rotational torques

In this section, we develop a theory of rotational torques on skyrmion crystals. Operating on the Landau-Lifshitz-Gilbert equation with $\frac{1}{|\gamma|} \hat{n} \times$ , where $\hat{n}$ is the direction of the magnetization and $\gamma$ is the gyromagnetic ratio, we obtain the equation of motion

$$m_0 \, \hat{n} \times \frac{d\hat{n}}{dt} = -\frac{\delta E}{\delta \hat{n}} - \alpha \, m_0 \frac{d\hat{n}}{dt} \qquad (S1)$$

where $m_0$ is the spin-density (with units of $\hbar$ per volume), and $\vec{b} = -\frac{\delta E}{\delta \hat{n}}$ is an effective magnetic field defined by the (functional) derivative of the total energy with respect to $\hat{n}$. A parametrizes the Gilbert damping.

We parametrize the magnetic texture by a rotation angle $\theta$ and write

$$\frac{d\hat{n}}{dt} = \frac{d\hat{n}}{d\theta} \partial_t \theta + \partial_t \hat{n}$$

with

$$\frac{d\hat{n}}{d\theta} = \hat{z} \times \hat{n} - (\hat{z} \times \vec{r}) \cdot \nabla \hat{n}. \tag{S2}$$

The first term describes the rotations of the spin orientation, the second one the rotation of space. To obtain an equation for the rate of change of the total angular moment, $J_z$, we multiply Eq. (S1) with $\frac{d\hat{n}}{d\theta}$ and integrate over space. After identifying the left-hand side of the equation with the time-derivative of the total angular momentum, we obtain

$$\frac{d}{dt} J_z = -\frac{dE}{d\theta} - \alpha \, m_0 \int d^3\vec{r} \, \frac{d\hat{n}}{d\theta} \left( \frac{d\hat{n}}{d\theta} \partial_t \theta + \partial_t \hat{n} \right). \tag{S3}$$

The total angular momentum, $J_z = S_z + L_z$, has two contributions [26], the first one, $S_z = m_0 \int d^3\vec{r} \ \hat{n}_z$, is simply the total spin in z-direction. The orbital angular momentum $L_z$ instead can be computed from the topological charge density $\rho_T = \frac{1}{4\pi} \hat{n} \cdot \left( \partial_x \hat{n} \times \partial_y \hat{n} \right)$. Up to surface terms, it can be written as [26], $L_z = -m_0 \int d^3\vec{r} \ 2\pi(x^2 + y^2) \rho_T(\vec{r})$.

To obtain an equation for the rotation angle, $\Delta\theta = \int dt \, \partial_t \theta$, after a field pulse, we simply integrate Eq. (S3) over time. Using that angular momentum is the same before and after the pulse, we obviously have $\int dt \frac{d}{dt} J_z = 0$. Furthermore, all rotationally invariant terms in the energy functional (exchange coupling, DMI interaction, dipolar interactions for spherical samples, magnetic fields in the z-direction) also do not contribute. In the limit of a weakly perturbed skyrmion crystal, we write the remaining terms as

$$\alpha \, D_{\text{rot}} \, \Delta\theta \approx \int dt \ \left( T_{\text{an}}^R + T_{\alpha,\text{pump}}^R + T_{\text{dis}}^R \right). \tag{S4}$$

The term on the left-hand side arises from the friction connected with the rotation of the skyrmion crystal with $D_{\text{rot}} = m_0 \int d^3 \, \vec{r} \left( \frac{d\hat{n}}{d\theta} \right)^2$ where we approximate $\hat{n}$ by the unperturbed spin texture $\hat{n}_0$.

The main contribution to this integral arises from the terms growing linear in $\vec{r}$ in Eq. (S2), giving rise to a contribution proportional to $r^2$ and thus linear in the number of skyrmions, $N_S$, involved in the rotation. Collecting those, we obtain for a roughly spherically shaped domain of a triangular skyrmion crystal

$$D_{\text{rot}} = m_0 \int d^3 \vec{r} \left(\frac{d\hat{n}_0}{d\theta}\right)^2 \approx N_S A_S \frac{m_0}{2\pi} \int d^3\vec{r} \ (\nabla\hat{n}_0)^2, \qquad (S5)$$

where $A_S$ is the area of the skyrmion unit cell. On the right-hand side of Eq. (S4) we collect different types of rotational torques. The first one, $T_{\text{an}}^R \propto -sin\, 6\,\theta$ arises from weak anisotropy terms which, in a clean system, fix the relative orientation of skyrmion crystal and crystalline lattice. For our experiment these terms can be completely neglected as $\Delta\theta$ is much smaller than $\frac{2\pi}{6}$ (i.e., there is no ratchet motion from minimum to minimum of $E(\theta)$) and furthermore $\Delta\theta$ is independent of $\theta$.

Remarkably, in the driven system, the Gilbert damping-term can give rise to rotational torques which *induce* a rotation with

$$T_{\alpha, \text{pump}}^R = -\alpha\, m_0 \int d^3\vec{r} \ \frac{d\hat{n}}{d\theta} \partial_t \hat{n} \qquad (S6)$$

This term, evaluated below, vanishes in thermal equilibrium but is generically finite in a driven system. Finally, the disorder in the system and boundary terms also induces torques, $T_{\text{dis}}^R$, which include pinning terms counteracting a rotation and pumping terms supporting rotations, see below.

Expanding the magnetic texture around its equilibrium configuration, $\hat{n} = \hat{n}_0 + \delta\hat{n}$, we observe that for the time-integrated torque, $\int T_{\alpha, \text{pump}}^R dt$, the contribution linear in $\delta\hat{n}$ vanishes, thus the leading term is quadratic in $\delta\hat{n}$. We now consider the response to a magnetic field pulse parallel to the external magnetic field $B_0$ (i.e., perpendicular to the surface). $\tau$ is the duration of the pulse and $\delta B$ its amplitude, such that $B(t) \approx B_0 + \delta B\,\tau\,\delta(t)$ describes a single pulse. Therefore, $\delta\hat{n}$ is linear in $\delta B\,\tau$. For the rotation angle after a single pulse we obtain in the absence of disorder

$$\Delta\theta_{\alpha,\text{pump}} \approx -\frac{2\pi}{N_S A_S} \frac{\int dt\, d^3\vec{r}\, \frac{d\delta\hat{n}}{d\theta} \partial_t \delta\hat{n}}{\int d^3\vec{r}\, (\nabla\hat{n}_0)^2} \approx \gamma_{\alpha,\text{pump}} \frac{1}{\alpha\, N_S} \left(\frac{\delta B}{B_0}\right)^2 \left(\frac{\tau}{T_0}\right)^2. \qquad (S7)$$

where we used the oscillation period $T_0$ of the breathing mode to obtain the dimensionless ratio $\frac{\tau}{T_0}$. The factor $\frac{1}{\alpha}$ reflects that the time integral is proportional to this factor as the life-time of the breathing mode is of the order of $T_0/\alpha$. The remaining prefactor $\gamma_{\alpha,\text{pump}}$ with the unit of an angle depends only weakly on all remaining system parameters. It encodes both how efficient the field pulse is in inducing collective oscillations of the magnetic textures and how efficient these oscillations induce rotational torques using non-linearities in $T^R_{\alpha,\text{pump}}$.

A simple two dimensional schematic animation is shown in the supplementary movie, where only a single breathing mode is taken into account. To obtain a numerical estimate for $\gamma_{\alpha,\text{pump}}$, we perform a micromagnetic simulation using mumax [34]. We use periodic boundary conditions, thus effectively simulating an infinitely large system. Thus, rotations are absent in the simulations but we compute directly $T^R_{\alpha,\text{pump}}$ and $\gamma_{\alpha,\text{pump}}$ from Eq. (S7). We simulate a slab with a thickness of 60 nm using a unit cell of $66\text{nm} \times 114\text{nm}$ discretized into cubes with a width of 2nm. Magnetic parameters of $Cu_2OSeO_3$ are used in the simulation. The saturation magnetization is $M_S = 1.044 \times 10^5 \text{Am}^{-1}$, the exchange parameter is $A = 3.547 \times 10^{-13} \text{Jm}^{-1}$, and the bulk DMI is $D = 7.43 \times 10^{-5} \text{Jm}^{-2}$. The skyrmion crystal is stabilized by a static field of 0.06T, and the skyrmion spacing is about 66nm, leading to the unit cell area of the skyrmion crystal $A_S = 3.772 \times 10^3 \text{nm}^2$. For these parameters, we find that a field pulse mainly excites two collective modes with frequencies 4.8GHz and 5.0GHz, see Fig. S1, roughly matching the experimentally observed frequency of 5.7Gz. As shown in the supplementary animation, both modes can be viewed as breathing modes with an oscillating magnetization amplitude but the modes differ by the z-dependence of their amplitude.

To make the simulations feasible, we use field pulses with a duration of 10ps, 200 times longer than the experimental pulses, and also relatively large values of the Gilbert damping $\alpha \sim 0.01$, but use Eq. (S7) to extrapolate to experimental values. Under a pulse of 10 mT, typical time variations of $N_S\dot{\theta}(t) \approx -\frac{2\pi}{A_S} \frac{\int d^3\vec{r} \frac{d\delta\hat{n}}{d\theta} \partial_t \delta\hat{n}}{\int d^3\vec{r} (\nabla \hat{n}_0)^2}$ and $N_S\theta(t) = \int_0^t N_S\dot{\theta}$ as are shown in Fig. S2. A finite rotation angle builds up on the time scale of a few ns for $\alpha = 0.01$. In Fig. S3 we show that the rotation angle is proportional to $(\delta B)^2/\alpha$ as predicted by Eq. (S7). Fitting the results to Eq. (S7), we obtain for a single pulse of square shape

$$\gamma_{\alpha, \text{pump}} \approx 9.7° \qquad (S8)$$

Thus, we have shown that Gilbert damping is able to induce sizable rotational torques but the experimental effects turns out to be much larger.

To compare to the experiment, we use that the pulse duration is $\Delta\tau = 50\,fs$ generating a field of about 13 mT via the inverse Faraday effect. Using that the oscillation period of the breathing mode is $175\,ps$, we estimate

$$\Delta\theta_{\alpha,\text{pump}} \approx \frac{1.1° \cdot 10^{-7}}{N_s\alpha} \ll 1° \cdot 10^{-6}$$

where we assume that $\alpha > 10^{-4}$ and that about hundred skyrmions rotate simultaneously in our sample. Clearly, the pumping induced by the Gilbert damping alone, cannot explain our experiment which shows rotation angles which are about 6 orders of magnitude larger.

Therefore, a different mechanism to generate rotational torques is needed to explain the experiment.

Above, we showed that the Gilbert damping $\alpha$ has two different roles. On one hand, it provides damping (left-hand side of Eq. (S4)), thus counteracting any rotation. But on the other hand, via $T_{\alpha,\,\text{pump}}^R$ the Gilbert damping also transforms the oscillatory breathing-mode into a rotational torques which induces rotations. The same also applies for disorder. First, it is the source of (collective) pinning by disorder, which counteracts any rotation. Second, in the presence of disorder there will be a ratchet-like motion where disorder induces rotations of the skyrmion crystal. A quantitative numerical simulation of this problem is challenging (and beyond the scope of this paper) as the rotational motion and the depinning physics is very complicated (involving defects in the skyrmion crystal [12]) and one must consider very large systems. Particle-based models as in Ref. [12] cannot be used as they do not include the physics of the breathing mode. Previous simulations of the motion of single skyrmions [28] in the presence of oscillating magnetic fields show that relatively fast motion can be induced along defect tracks. Thus, we conclude that defect-assisted rotational torques (perhaps together with surface terms) induced by breathing-mode oscillations are likely to explain our experiment.

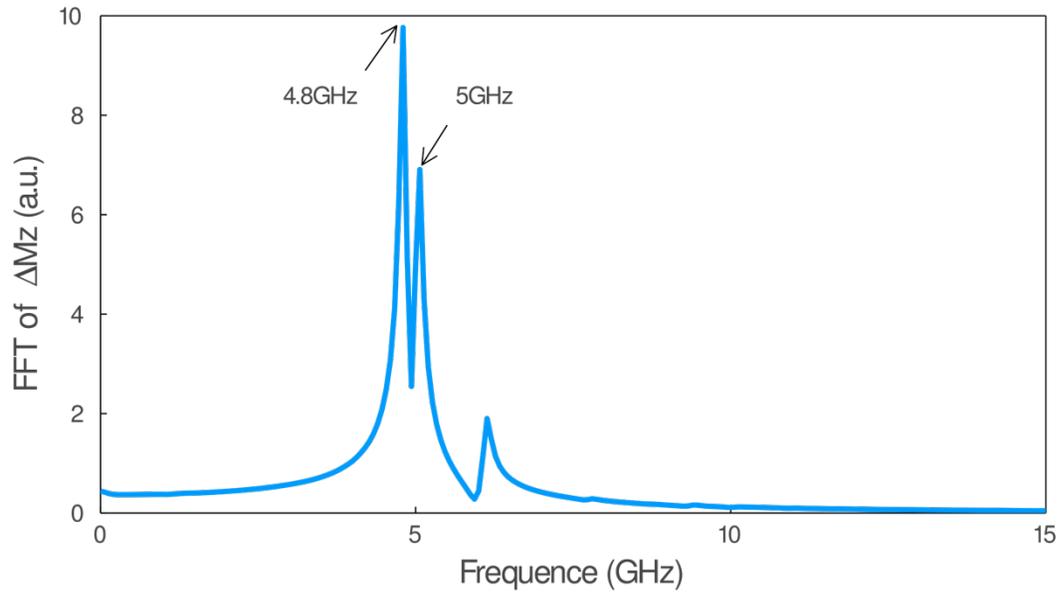

FIG. S1: Fast Fourier transformation of the total excited magnetization as a function of the excitation frequency obtained for α = 0.01. Two collective breathing modes are labeled.

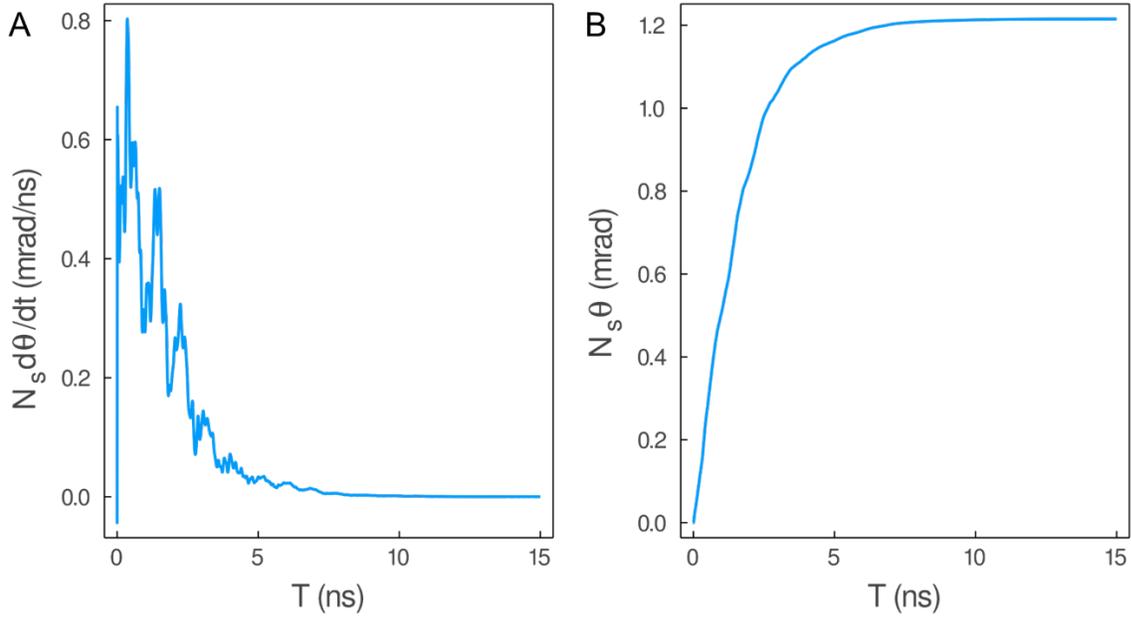

FIG. S2: Typical time variations of (a) the rotation rate and (b) total rotation angle $N_S\theta(t) = \int_0^t N_S\dot{\theta}$ for $\alpha = 0.01$, for a square-shaped pulse with a duration of 10ps and an amplitude of 10mT. The rotation builds up on a time scale of a few ns for these parameters.

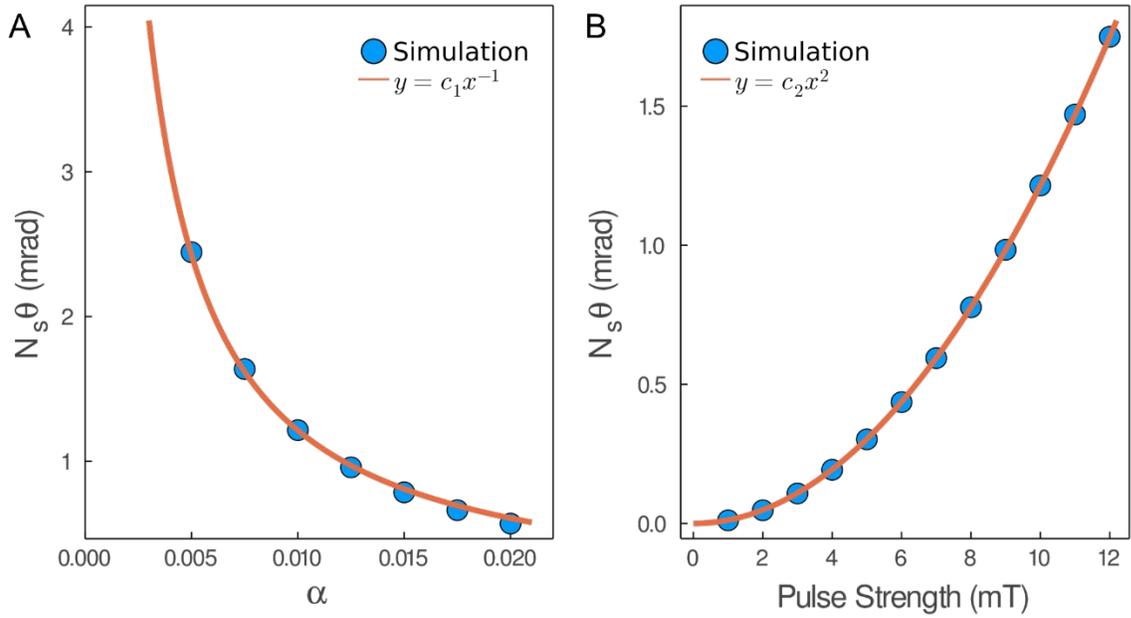

FIG. S3: As predicted by Eq. (S7), the rotation angle is inversely proportional to the Gilbert damping α and quadratic in the pulse strength (parameters: pulse strength of 10mT for panel (a) and α = 0.01 for panel (b), pulse duration 10ps).

2. Estimation of inverse Faraday effect by pump beam

The formula we use is[35]:

$$M = \frac{\lambda V}{2\pi c}(I_R - I_L) \tag{S9}$$

Where $I_R$ is the peak intensity of the right hand circularly polarized pulse. These values are in the cgs system.

$$V = 3.0 \; x \; 10^{-2} \; \frac{rad}{Oe \; cm}$$

The experimental parameters:

$$\lambda = 1.2 \; x \; 10^{-4} \; cm$$

$$E = 0.2 \frac{uJ}{pulse} = 2 \frac{erg}{pulse}$$

$$\tau = 50 \; fs$$

$$d = 40 \; um$$

$$I_{R,peak} = 2x \frac{2}{(50x10^{-15})\pi(20x10^{-4})^2} = 6.57x10^{18}$$

$$M = \frac{(1.2x10^{-4})(3x10^{-2})}{2\pi(3x10^{10})}(6.57x10^{18}) = 121.6 \; G = 12.2 \; mT$$

To calculate the effective magnetic field $B_{eff}$ resulting from the pulse, we used M/H values measured in Ref [36]:

$$M = 0.1 \; \mu_B/Cu^{2+} at \; H = 30mT$$

Given the concentration of Cu atoms $= \frac{16}{8.911\text{Å}} = 2.26 x 10^{28} m^{-3}$

We find $H_{eff} = 12.1$ mT $\left(\frac{30mT}{26.3mT}\right) = 14$ mT

### 3. Details on the gridded analysis of skyrmion rotation in our image

The full data set used to generate Figure 4 in the main text is shown in Figure S4 below. The real space image was divided into regions of 460 nm. From each gridded subsection, the Fourier Transform was taken, and a coordinate transform is made from cartesian to polar coordinates. We then integrate over a certain region in the radial coordinate. Due to symmetry, we average the date from angles 0-180° with data from angles 180-360°. We then plot intensity vs. angle and choose the peak with the highest intensity and plot a region in angle space around this peak. This provides us with intensity maps in the angle coordinate. By fitting these intensity maps with a gaussian function, we extract the central peak and use this to plot the function of angle vs. number of pulses seen in Fig. 1 of the main text, and to calculate the fluence dependence for Fig. 2a of the main text.

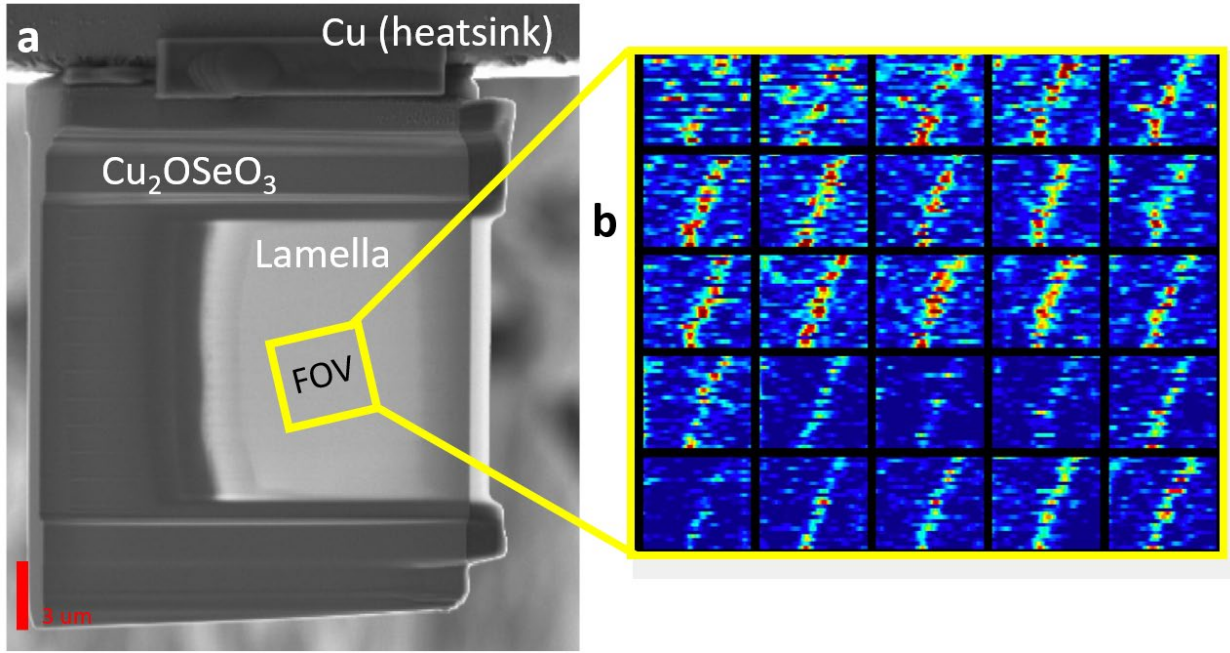

FIG. S4. (a) Scanning electron microscope image of lamella. The scale bar (bottom left) is 3 um. The range of interest (ROI) of our L-TEM images is given by the yellow box. (b) Each box maps the angle of a single point in the FT of a subsection of the image following excitation during a train of circularly polarized laser pulses (1 image for each pulse. The number of pulses sent on the y axis was 16 and the x-axis runs from 3.5 to 88 degrees.

5. Finite Element modelling of timescales and final temperature reached following laser pulse excitation

To understand and rule out all thermal timescales relevant for the rotation process, it is important to identify the evolution of temperature in the sample following the absorption of the laser pulse. We modeled the heating and cooling dynamics of the sample following ultrafast laser excitation using the finite element method (See SI for further details). Our model uses experimental values for the heat capacity and heat conduction in $Cu_2OSeO_3$ [9], [37]. At the threshold fluence (2.2 mJ/cm$^2$), our calculations show that the sample reaches a maximum temperature of ~27K and remains at that temperature for at least 2-3 ns, followed by cooling via the cold finger on the timescale of 100s of nanoseconds. We then performed the simulation with two laser pulses, separated by a delay of 100 ps (as was done in the experiment and resulted in no rotation, see Fig. 2b). We find that the peak temperature reached does not change significantly despite the separation of the two pulses by 100 ps. This information combined with the fact that the cooling timescale (100s of ns) is much slower than the dynamics that we observe in our time resolved measurements and the coherent nature of the rotations observed, all serve to show that thermal effects are not relevant for the rotation dynamics observed here.

The thermodynamic finite element model of the $Cu_2OSeO_3$ sample, heated by an ultrafast laser pulse, has been implemented using *COMSOL*. The geometry implemented is shown in Fig. S6. Here, the domain is divided in two regions: the lamella (in red), with a thickness of 100 nm, and the support (green), with a thickness of $1\ \mu m$.

The experimental sequence has been described analytically as a two pulse Gaussian beam, having a flux $\phi(x, y, t) = T_{opt} \cdot (1 - R)\, \phi_{inc} \cdot exp(-2\, r_f{}^2/r_s{}^2) \cdot (gp(t, \tau_{pulse}, t_1) + gp(t, \tau_{pulse}, t_2))$, with a spot radius of $r_s = 20\ \mu m$ and incident from the top with a power density $\phi_{inc} = 2.2 \times 10^{10}\ W/cm^2$ (corresponding to the incident fluence of $1.1\ mJ/cm^2$) in each pulse. $r_c$ is the radial distance from the beam axial center and is defined as $r_c{}^2 = (x - x_c)^2 + (y - y_c)^2$, with $x_c$ and $y_c$ being the planar coordinates of the central axis. $T_{opt} = 0.88$ is the transmission coefficient of the optical system (window and reflective mirror inside the microscope) while $R = 0.11$ is the reflection coefficient of $Cu_2OSeO_3$ at $\lambda = 1.24\ \mu m$ (1 eV). $gp(t, \tau_{pulse})$ is a normalized gaussian pulse with duration $\tau_{pulse} = 50\ fs$, centered around $t_1 = 0.5\ ps$ for the first pulse and $t_2 = t_1 + 100\ ps$ for the second one, to model the first experimentally observed null (no rotation) in the double pulse experiment (see

Figure 2b of main text). The heating induced on the sample by the laser pulse has been introduced through a volumetric heat source, modelled as $Q_0 = \alpha \, \phi_{inc} \, exp(\alpha(z - z_0))$, where $\alpha = 318.66 \, cm^{-1}$. Here, $z_0$ corresponds to the longitudinal coordinate of the top surfaces for both lamella and support. One of the support's lateral surfaces has been fixed to the temperature of 5K, emulating the behavior of the heatsink. Initial temperature has been set to 5K all over the structure.

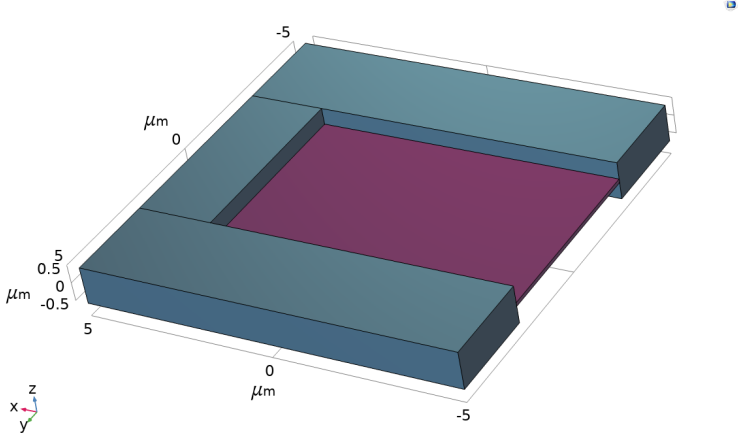

FIG. S5. Geometry of the Cu2OSeO3 sample implemented in Comsol Multiphysics. The domain has been divided in two regions: the lamella in purple and the support in grey.

The time dependent simulation is run for a period of $1 \, \mu s$, considering the temperature dependent heat capacity $C(T)$ and heat conductivity $k_L(T)$ discussed in Section 4, with non-uniform time steps.

The temperature trend resulting from the simulation are shown in Fig. S7 in case two pulses, spaced $100 \, ps$ apart, heat the sample. Here, we show the maximum, minimum and average temperatures reached by the sample as a whole (Fig. S10A) and by the lamella (Fig. S10B) following the laser excitation. Fig. S10A shows that the absolute maximum temperature reached by the support are of 21.68 K after the first laser pulse and 26.54 K after the second laser pulse. The average temperature slowly decays to avg(T) ~ 20 K reached in $t \sim 6.8 \, ns$ and finally to 5K in a time interval of $0.5 \, \mu s$. Maximum temperature in the support is kept almost constant up to $t \sim 60 \, ns$, corresponding to the onset of the cooling of the last part of the support, the farthest from the heatsink. The minimum temperature measured all over the support is always 5K, representative of the heatsink surface. Fig. S10B shows that the absolute maximum temperature reached by the thin lamella is of 21.87 K after the first pulse and of 26.80 after the second. The temperature in the lamella is quite uniform during the heating phase, in fact max(T) – min(T) = 1-2 K. At $t \sim 1.5 \, ns$ the minimum temperature drops from ~25.6K to ~ 8K in 25 ns, meaning that the cold region, with the temperature imposed by the heatsink, is entering the lamella's domain.

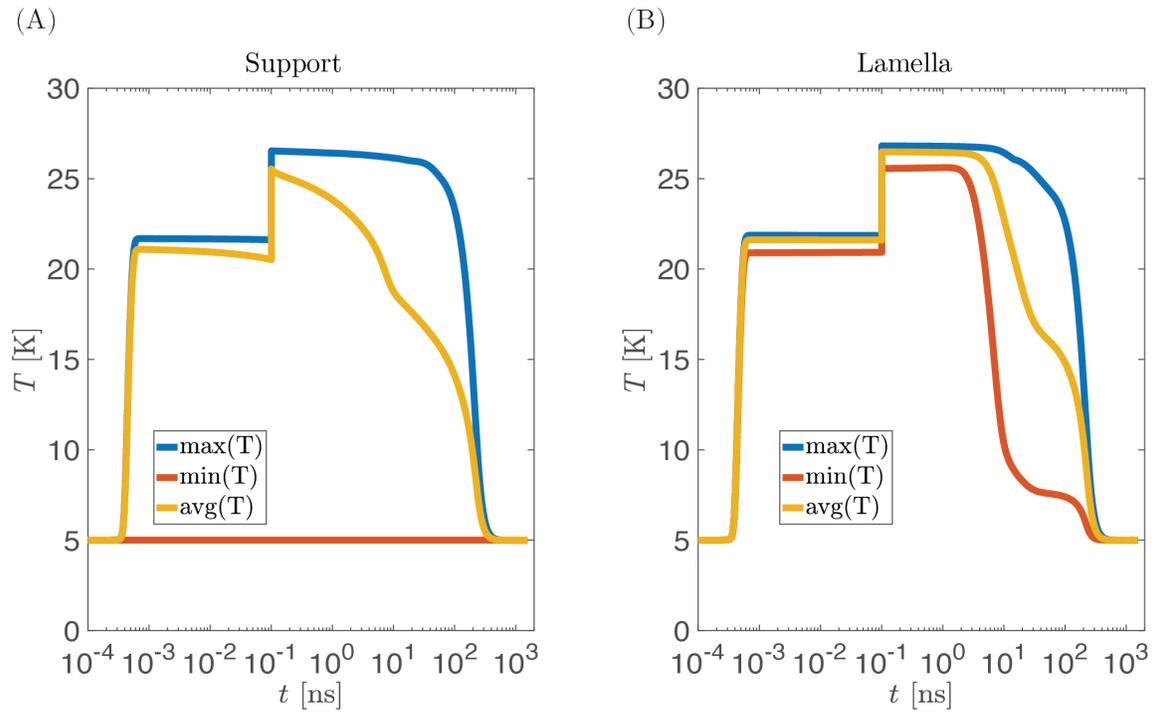

FIG. S6. Maximum (blue), minimum (red) and average (yellow) temperature obtained in the support (A) and in the lamella (B) in case two laser pulses, spaced 100 ps apart, heat the sample .

The entire process of cooling takes approximately $0.5 \, \mu s$.

Fig. S11, shows the isothermal contours obtained on the top surface at $t = 120\ ps$, thus $20\ ps$ after the second pulse hit the sample. The gradient imprinted on the surface is circular due to the gaussian shape of the laser beam. Note that within the region of the lamella (where measurements are taken), the heating is nearly uniform (gradient less than 1 K over 10 um).

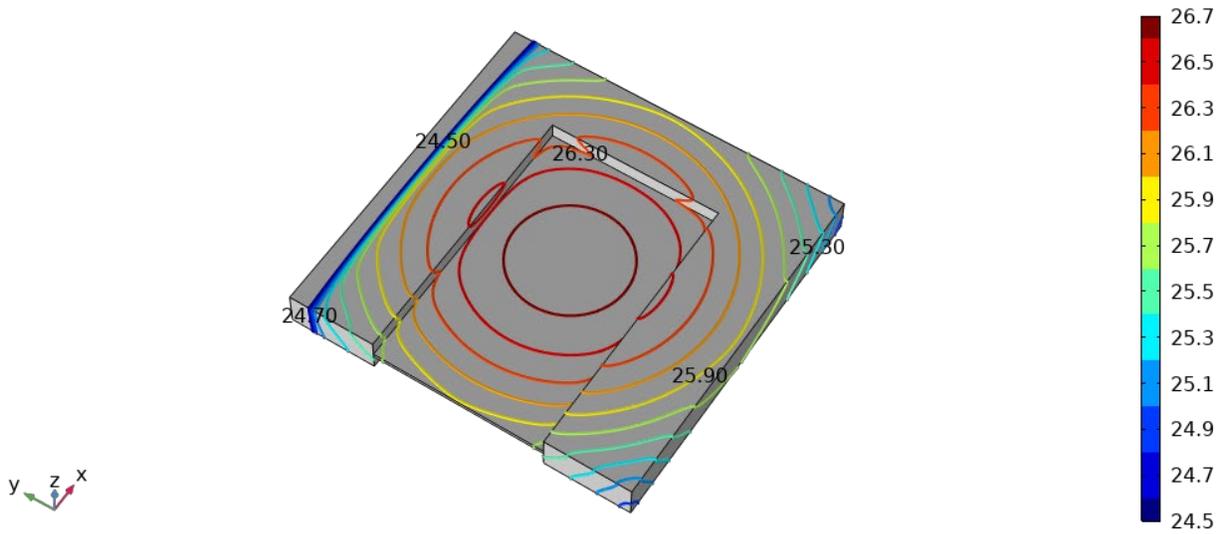

FIG. S7. Isothermal contours after two pulses laser irradiation ($t = 120\ ps$) on the top surface. Black labels represent the temperature. Heat sink surface is on the top left.

Next we show the case when the two $1.1\ mJ/cm^2$ pulses arrive simultaneously, i.e. $t_2 = t_1$. Fig. S12 shows the temperature evolution. In this case, the maximum temperature reached in the lamella is of 25.81 K, while the one in the support is of 26.57 K, which is very similar to the case of Fig. S10.

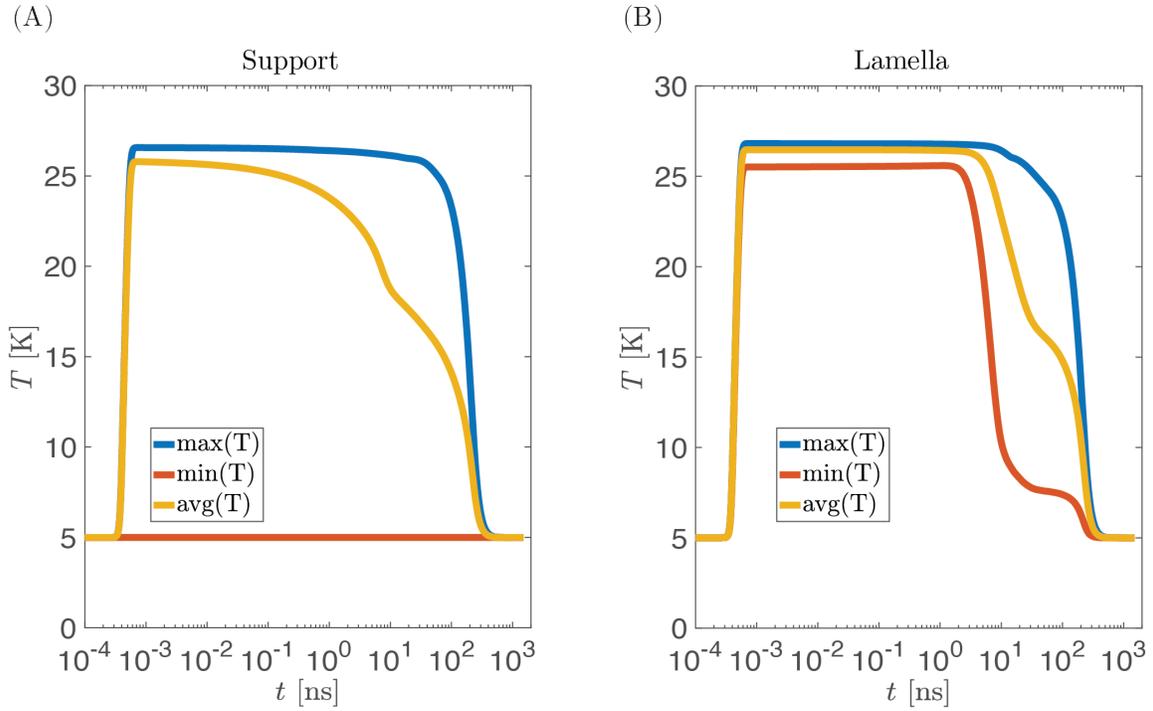

FIG. S8. Maximum (blue), minimum (red) and average (yellow) temperature obtained in the support (A) and in the lamella (B) in case two laser pulses arrive simultaneously.

Fig. S13 shows the temperature map at $t = 30\ ns$ and at $t = 90\ ns$, thus during the cooling process of the lamella. The gradient, during this phase, is imprinted by the heatsink, leading to a linear boundary between the hot and cold domains.

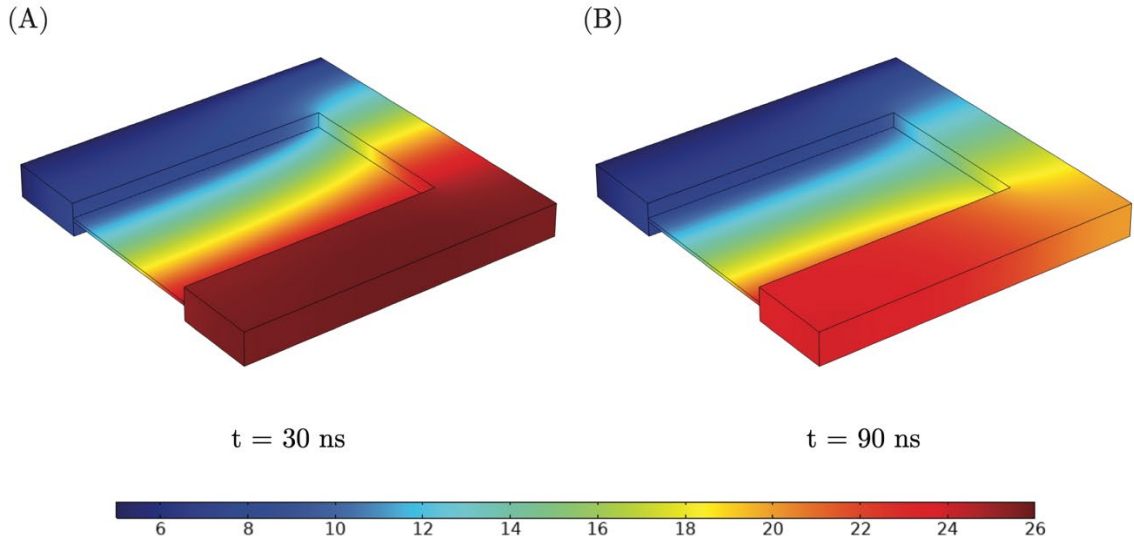

<div align="center">

(A)          (B)

t = 30 ns          t = 90 ns

</div>

FIG. S9. Temperature map of the Cu2OSeO3 sample during the cooling phase in two different time instants: at $t = 30\ ns$ (A) and $t = 90\ ns$ (B). Color legend indicates the temperature in Kelvin. The top left surface is representative of the heat sink and it is kept at the constant temperature of 5K during the simulation.

The peak temperature of ~27 K reached in the film is below the paramagnetic melting point of the skyrmion crystal, thus the sample can still "remember" the previous skyrmion orientation, which explains well the rotation observed in our images. This is a fundamentally different mechanism than many previous studies of skyrmion or magnetic bubble formation[38], [39].

6: Details of the phononic heat capacity and heat conductivity used in FEM

Experimental data on the temperature dependent heat capacity of $Cu_2OSeO_3$ has been retrieved from Ref. [40]. Fitting these data with the following analytical model for heat capacity we retrieved a Debye Temperature $\theta_D$ of 287 K:

$$C_{ph}(T) = A\left(\frac{T}{\theta_D}\right)^3 \int_0^{\theta_D/T} \frac{x^4}{(e^x - 1)(1 - e^{-x})}\, dx \qquad (S10)$$

where A and $\theta_D$ are fitting parameters.

For the heat conductivity we used the Callaway model [41], together with its update [42], as done in Ref. [9]. In particular, the integral expression for the phononic heat conductivity is the following:

$$k_L = \frac{k_B}{2\pi^2 v}\left(\frac{k_B}{\hbar}\right)^3 T^3 \left[\int_0^{\theta_D/T} \frac{x^4 e^x}{(e^x - 1)^2}\, \tau_c(x,T)\, dx\right]$$
$$\times \left(1 + \frac{\overline{\tau_c(T)/\tau_N(T)}}{\tau_c(T)/\tau_R(T)}\right) \qquad (S11)$$

with

$$\frac{\overline{\tau_c(T)/\tau_N(T)}}{\tau_c(T)/\tau_R(T)} = \int_0^{\theta_D/T} \frac{x^4 e^x}{(e^x - 1)^2}\, \frac{\tau_c(x,T)}{\tau_N(x,T)}\, dx \left/ \int_0^{\theta_D/T} \frac{x^4 e^x}{(e^x - 1)^2}\, \frac{\tau_c(x,T)}{\tau_R(x,T)}\, dx \right. \qquad (S12)$$

where $v$ is the Debye-averaged sound speed, $x = \hbar\omega/k_BT$ is the reduced phonon energy and $\tau_C^{-1}(x,T) = \tau_N^{-1}(x,T) + \tau_R^{-1}(x,T)$. In the latter equation $\tau_N^{-1}(x,T)$ and $\tau_R^{-1}(x,T)$ are the phonon scattering rates for the normal (that conserve momentum) and resistive processes (that do not conserve the momentum), respectively. Their expressions are given below:

$$\tau_R^{-1}(x,T) = v/l_{ph} + A\, x^2 T^4 exp\left(-\frac{\theta_D}{b\, T}\right) + C x^4 T^4 \qquad (S13)$$

$$\tau_N^{-1}(x,T) = \gamma A \, x^2 \, T^4 \qquad\qquad \text{(S14)}$$

$\tau_R^{-1}(x,T)$ includes terms accounting for scattering from boundaries (Umklapp scattering), other phonons and point-like defects. Here A, b, C and $\gamma$ are fitting parameters. To model our 10 um Cu₂OSeO₃ sample, we used for A, b, C and $\gamma$ the values reported in Ref. [9] for the smallest available $k_L$ ($l_0 = 0.31 \, mm$), revaluating $k_L$ assuming $l_{ph} = l_0 = 10$ um. Fig. S15 reports the heat conductivity obtained following the aforementioned model. More precisely the fitting parameters used are: $A = 1.5 \times 10^4 K^{-4}, b = 6.35, \ C = 110 \, K^{-4}, \ $ and $\gamma = 0. \ \theta_D$ has been chosen equal to 287 K to be coherent with the heat capacity model. The condition $\gamma = 0$ is equivalent to set $\overline{\dfrac{\tau_c(T)/\tau_N(T)}{\tau_c(T)/\tau_R(T)}} = 0$.

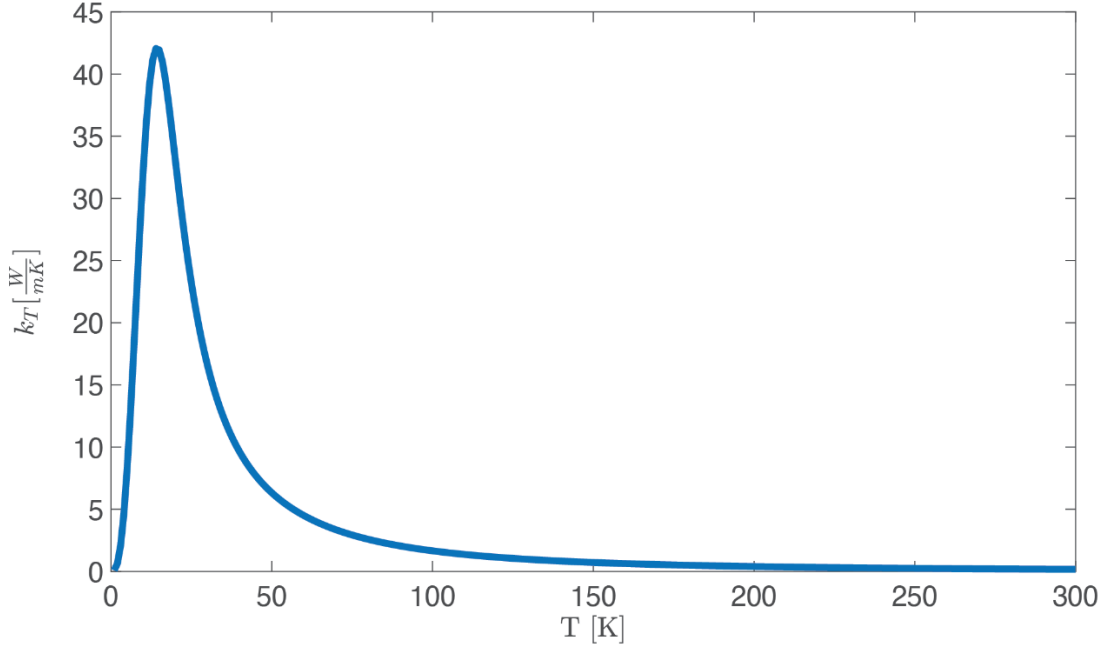

FIG. S10. Analytical heat conductivity obtained with the Callaway model using the fitting parameters of Ref. [13], $\theta_D = 287$ K and $l_{ph} = l_0 = 10$ um.